\documentclass[aps,prl,onecolumn,groupedaddress,superscriptaddress,notitlepage]{revtex4-1}
\usepackage{color,graphicx}
\usepackage{subfigure}
\usepackage{amsmath}
\usepackage{amsthm}
\usepackage{float}
\usepackage{epstopdf}
\newcommand{\ra}{\rangle}
\newcommand{\la}{\langle}
\newcommand{\bra}[1]{\ensuremath{\langle #1|}}
\newcommand{\ket}[1]{\ensuremath{|#1\rangle}}

\begin{document}
\title{Non-Markovian Dynamics of Macroscopic Quantum Systems in Interaction with Non-Equilibrium Environments}
\author{Nasim Shahmansoori}
\affiliation{Research Group On Foundations of Quantum Theory and Information, Department of Chemistry, Sharif University of Technology, P.O.Box 11365-9516, Tehran, Iran}
\author{Farhad Taher Ghahramani}
\affiliation{Foundations of Physics Group, School of Physics, Institute for Research in Fundamental Sciences (IPM), P.O.Box 19395-5531, Tehran, Iran}
\author{Afshin Shafiee}
\email[Corresponding Author:~]{shafiee@sharif.edu}
\affiliation{Research Group On Foundations of Quantum Theory and Information, Department of Chemistry, Sharif University of Technology, P.O.Box 11365-9516, Tehran, Iran}
\affiliation{Foundations of Physics Group, School of Physics, Institute for Research in Fundamental Sciences (IPM), P.O.Box 19395-5531, Tehran, Iran}

\begin{abstract}
We study the dynamics of a macroscopic superconducting qubit coupled to two independent non-stationary reservoirs by using time-dependent perturbation theory. We show that an equilibrium environment surpasses the coherent evolution of the macroscopic qubit completely. When the qubit couples to two different reservoirs, exemplifying a non-equilibrium environment, the short-time dynamics is affected by the interference between two reservoirs, implying the non-additivity of effects of two reservoirs. The non-additivity can be traced back to a non-Markovian effect, even though two reservoirs are independently assumed to be Markovian. Explicitly, the non-equilibrium environment intensifies both coherent and incoherent parts of the evolution. Therefore, the macroscopic qubit would evolve more coherently but at the price of a shorter decoherence time.
\end{abstract}

\maketitle
\section{1. Introduction}
An environment in thermal equilibrium destroys the coherence of a quantum system interacting with it~\cite{Joo,Schl,Bre,Wie}. Such decoherence process is one of the main obstacles in realizing quantum computers~\cite{DiV,Unr,Sho,Pal}. One approach to implement a quantum computer is based on superconductors, where the quantum effects become macroscopic, though at a price of extremely low temperatures~\cite{Dev,Mot,Cho,Kel}. Apart from this straightforward strategy, a number of schemes have been proposed to control the decoherence process by engineering the system-environment interaction (for a rather complete review, see~\cite{Sut}). One of the effective strategies, observed naturally in biological systems~\cite{Chi,Hue}, is engineering non-equilibrium environments. Such an environment has the opportunity to influence the quantum evolution in a manner that is more rich and complex than simply acting to randomize relative phases and dissipate energy.\\
\indent The open quantum systems are mostly examined by Lindblad master equations~\cite{Lin,GorK,Bre}, which are based on Born and Markov approximations, and depending on the context, an additional uncontrolled approximation~\cite{Bre,Riv,Gur,Ste} on the environment. However, in non-equilibrium systems, such approximations may lead to incorrect predictions~\cite{Wic,RiP,Mar,Lev,Bar,Sto,Str}. One of such predictions is that the effect of two initially uncorrelated environments on the system's dynamics is independent and additive in the weak-coupling limit~\cite{Scha}. The validity of this prediction has been examined in a number of contexts in the literature~\cite{RiP,Lev,Sto,Sca,Pur,Tru,Eas,Dec,Hof,Gon,Mit}. In fact, the non-additive effect is not unfolded by Markovian approach, implying the non-Markovian property of the dynamics~\cite{Rib,Hal,Pet,Fer}. \\
\indent Apart from the approach employed, among all the works mentioned, the lack of a macroscopic quantum system as the case study is deeply felt. Here, we examine the effective dynamics of a macroscopic superconducting qubit, in interaction with a non-equilibrium environment. To be macroscopic, conceptually, the dynamics should involve macroscopically distinguishable, entangled states~\cite{Sek}. The effective dynamics of the macroscopic superconducting qubit, particularly the phenomenon of macroscopic quantum coherence, can be studied in the typical double-well potential~\cite{Tak}. At sufficiently low temperatures, the Hamiltonian of the system can be expanded by two first states of energy, each of which, as states of a macroscopic system, involves a large number of bosonic particles~\cite{Leg}. The experiments devoted to the studies of macroscopic quantum coherence - most notably, those involving superconductivity, superfluidity, and single-domain magnet - require temperatures close to absolute zero to operate (for more detail, see~\cite{Tak}, ch.3). The non-equilibrium environment can be engineered as two independent vacuum reservoirs with different spectral densities. To address non-equilibrium effects, we employ the time-dependent second-order perturbation theory. We demonstrate that the resulting environmental effect is non-additive. \\
\indent The paper is organized as follows. In the next section, the standard form of Hamiltonians is presented. In section 3, we describe the physical model of the total system, consisting of the macroscopic quantum system and surrounding reservoirs. We then examine the kinematics and then the dynamics of the system in sections 4 and 5, respectively. The parameters of the model for a macroscopic superconducting qubit as the central system are estimated in section 6. The results are discussed in section 7. Our concluding remarks are presented in the last section.
\section{2. Standard Form of Hamiltonian}
To quantify the macroscopicity of the system, we incorporate the dimensionless form of the model. The macroscopic system, composed of a large number of bosonic particles, oscillates in time between macroscopically distinct states. Such a system is typically modeled by the motion of a quasi-particle of mass $\mathsf{M}$ in a symmetric double-well potential (see section 6) with Hamiltonian
\begin{equation}\label{Hs}
\mathsf{H}_{\mathsf{S}}=\frac{\mathsf{P}^{2}}{2\mathsf{M}}+\mathsf{U}(\mathsf{R}),
\end{equation}
The potential $\mathsf{U}(\mathsf{R})$ can be represented by a quartic function as
\begin{equation}\label{U}
\mathsf{U}(\mathsf{R})=\frac{\mathsf{U}_{0}}{2}\Big(\frac{\mathsf{R}^{2}}{\mathsf{R}_{0}^{2}}-1\Big)^{2};~~~ \mathsf{U}_{0}=\frac{\mathsf{M}\mathsf{\Omega}^{2}\mathsf{R}_{0}^{2}}{4},
\end{equation}
where $\mathsf{\Omega}$ is the harmonic frequency at the bottom of each well, and $\mathsf{R}_{0}$ is the half distance between two minima. Here, the strength of the potential is parametrized such that $\frac{\partial^{2}\mathsf{U}}{\partial \mathsf{R}^{2}}|_{\mathsf{R}=\mathsf{R}_{0}}=\mathsf{M}\mathsf{\Omega}^{2}$, just as the standard harmonic potential. The potential has the characteristic energy $\mathsf{U}_{0}$ and the characteristic length $\mathsf{R}_{0}$, which we adopt as the units of energy and length. The corresponding characteristic time can be defined as $\mathsf{T}_{0}=\mathsf{R}_{0}/(\mathsf{U}_{0}/\mathsf{M})^{1/2}$, which we consider as the unit of time. Likewise, the unit of momentum is taken as $\mathsf{P}_{0}=(\mathsf{M}\mathsf{U}_{0})^{1/2}$. We define the dynamical variables, $x$ and $p$, as $\mathsf{R}/\mathsf{R}_{0}$ and $\mathsf{P}/\mathsf{P}_{0}$. The corresponding commutation relation is defined as $[x,p]=ih$, where the Planck constant is redefined as $h=\hbar/\mathsf{R}_{0}\mathsf{P}_{0}=\hbar/\mathsf{U}_{0}\mathsf{T}_{0}$. We consider the reduced Planck constant ``$h$'' as the measure of macroscopicity of the system of interest. In section 6, we calculate the parameter $h$ for a macroscopic superconducting qubit. Note that we denote dimensional variables with sans serif font and dimensionless ones by italic fonts. \\
\section{3. Model}
The Hamiltonian of the total system, composed of the macroscopic quantum system $S$ and the reservoirs, $A$ and $B$, is conveniently defined as
\begin{equation}\label{1}
H=H_{S}+\sum_{\mathcal{R}}H_{\mathcal{R}}+\sum_{\mathcal{R}}H_{S\mathcal{R}},
\end{equation}
where $\mathcal{R}=A,B$. The Hamiltonian of the system in the dimensionless form would be
\begin{equation}\label{HsD}
H_{S}=\frac{p^{2}}{2}+\frac{\Omega^{2}}{4}(x^{2}-1)^{2},
\end{equation}
where we define $\Omega^{2}\equiv8$. In the limit $k_{\mbox{\tiny$B$}}\mathsf{T}\ll \mathsf{U}_{0}$ (with $k_{\mbox{\tiny$B$}}$ as Boltzmann constant, and $\mathsf{T}$ as temperature), the states of the system are confined in the two-dimensional Hilbert space spanned by the lowest localized eigenstates of right and left wells, denoted by $|L\rangle$ and $|R\rangle$~\cite{Aha}. Hence, the effective Hamiltonian of the system in the localized basis is
\begin{equation}
H_{S}=-\Delta h\big(|L\rangle\langle R|+|R\rangle\langle L|\big),
\end{equation}
where $\Delta$ is the tunneling strength between two localized states. For an isolated system, the probability of the tunneling from the left localized state to the right one is given by
\begin{equation}\label{PI}
P_{L\rightarrow R}=\sin^{2}\big(\frac{\Delta t}{2}\big).
\end{equation}
\indent A frequently employed model for a reservoir is a collection of harmonic oscillators~\cite{KL}. The $\alpha$-th harmonic oscillator in the reservoir $\mathcal{R}$ is characterized by its natural frequency, $\omega_{\alpha,\mathcal{R}}$, and position and momentum operators, $x_{\alpha,\mathcal{R}}$ and, $p_{\alpha,\mathcal{R}}$, respectively, according to the Hamiltonian
\begin{equation}\label{HE}
H_{\mathcal{R}}=\sum_{\alpha}\frac{1}{2}\Big(p_{\alpha,\mathcal{R}}^{2}+\omega_{\alpha,\mathcal{R}}^{2}x_{\alpha,\mathcal{R}}^{2}-h\omega_{\alpha,\mathcal{R}}\Big).
\end{equation}
The last term which merely displaces the origin of energy is introduced for later convenience. For the reservoir $\mathcal{R}$, we define $|0\rangle_{\mathcal{R}}$ as the vacuum eigenstate and $|\alpha\rangle_{\mathcal{R}}$ as the single-boson exited eigenstate with energy $E_{\alpha,\mathcal{R}}$.\\
\indent The interaction between the macroscopic quantum system and the reservoir $\mathcal{R}$ has the form~\cite{Tak}
\begin{equation}\label{Hint}
H_{S\mathcal{R}}=-\sum_{\alpha}\Big(\omega_{\alpha,\mathcal{R}}^{2}f_{\alpha,\mathcal{R}}(x)x_{\alpha,\mathcal{R}}+\frac{1}{2}\omega_{\alpha,\mathcal{R}}^{2}f_{\alpha,\mathcal{R}}^{2}(x)\Big),
\end{equation}
according to which the macroscopic quantum system displaces the origin of the oscillator $\alpha$ of reservoir $\mathcal{R}$ with the spring constant $\omega_{\alpha,\mathcal{R}}^{2}$ by $f_{\alpha,\mathcal{R}}(x)$, as it can be recognized from
\begin{equation}
H_{\mathcal{R}}+H_{S\mathcal{R}}=\sum_{\alpha}\frac{1}{2}\Big[p_{\alpha,\mathcal{R}}^{2}+\omega_{\alpha,\mathcal{R}}^{2}
\Big(x_{\alpha,\mathcal{R}}-f_{\alpha,\mathcal{R}}(x)\Big)^{2}-h\omega_{\alpha,\mathcal{R}}\Big].
\end{equation}
The second term of (\ref{Hint}), which depends only on the system coordinate, renormalizes the potential and provides invariance of the
coupling under spatial translation. The physical consequences of such renormalization have been addressed in a number of contexts. Caldeira and Leggett have demonstrated that one can expect a renormalization effect when a collective degree of freedom is coupled to many single-particle modes~\cite{KL}. Petruccione and Vacchini examined the necessity of such term to obtain a translationally-invariant reduced dynamics for the Brownian particle in a homogeneous fluid~\cite{Petr}. In our case, since a macroscopic superconducting qubit is characterized by a collective degree of freedom, the renormalization term is relevant. For simplicity, we assume that the interaction model is separable ($f_{\alpha,\mathcal{R}}(x) =\gamma_{\alpha,\mathcal{R}}f(x)$, where $\gamma_{\alpha,\mathcal{R}}$ is the coupling strength) and bilinear ($f(x)=x$).
\section{4. Stationary Perturbation Theory}
Let us begin by reviewing the standard stationary perturbation theory for our system. The shift in the energy of the system, $E_{n}$, due to the perturbation $H_{S\mathcal{R}}$ up to the second order is obtained as
\begin{align}\label{DE}
\delta E_{n,\mathcal{R}}&\simeq{}_{\mathcal{R}}\la0|\la n|H_{S\mathcal{R}}|n\ra|0\ra_{\mathcal{R}}+\sum_{m\neq n}\frac{\big|{}_{\mathcal{R}}\la\alpha|\la m|H_{S\mathcal{R}}|n\ra|0\ra_{\mathcal{R}}\big|^{2}}{E_{n}-(E_{m}+E_{\alpha,\mathcal{R}})}\nonumber\\
&=\frac{1}{2}\sum_{r}x_{rn}^{2}\Omega_{rn}\sum_{\alpha}\frac{\gamma_{\alpha,\mathcal{R}}^{2}\omega_{\alpha,\mathcal{R}}^{2}}{\omega_{\alpha\mathcal{R}}+\Omega_{rn}},
\end{align}
where $x_{mn}=\langle m|x|n\rangle$ and $\Omega_{mn}=E_{m}-E_{n}/h$. The state with the energy shifted to $E_{n}+\delta E_{n,\mathcal{R}}$ due to the perturbation is not actually stationary, rather it decays with a finite lifetime $\Gamma_{n,\mathcal{R}}^{-1}$, given by the Fermi's golden rule as
\begin{align}\label{GnR}
\Gamma_{n,\mathcal{R}}&\simeq\frac{2\pi}{h}\sum_{m\neq n}\big|{}_{\mathcal{R}}\la0|\la n|H_{S\mathcal{R}}|n\ra|0\ra_{\mathcal{R}}\big|^{2}\delta(E_{n}-(E_{m}+E_{\alpha,\mathcal{R}}))  \nonumber\\
&=\frac{\pi}{h}\sum_{r}x_{rn}^{2}\Omega_{rn}\sum_{\alpha}\gamma_{\alpha,\mathcal{R}}^{2}\omega_{\alpha,\mathcal{R}}^{2}\delta(\Omega_{rn}-\omega_{\alpha,\mathcal{R}}).
\end{align}
With mere knowledge of $\delta E_{n\mathcal{R}}$ and $\Gamma_{n,\mathcal{R}}$ , given the initial state of the left-handed state for the system and vacuum states for the reservoirs, the approximate total state at time $t$ would be
\begin{align}\label{kin}
|\Psi(t)&\simeq\big(1-\frac{1}{2}e^{-t/(\Gamma_{2,A}+\Gamma_{2,B})}\big)^{1/2}e^{-i(E_{1}+\delta E_{1,A}+\delta E_{1,B})t/h}|1\rangle|0\rangle_{A}|0\rangle_{B}\nonumber\\&-\frac{1}{2}e^{-t/2(\Gamma_{2,A}+\Gamma_{2,B})}e^{-i(E_{2}+\delta E_{2,A}+\delta E_{2,B})t/h}|2\rangle|0\rangle_{A}|0\rangle_{B}.
\end{align}
That is, the interaction excites only the system, and the environmental oscillators remain in the vacuum state (Born approximation). This is, however, inaccurate. In fact, the interaction with a macroscopic quantum system may excite the environmental oscillators too. So, it is necessary to examine the time evolution of the total system in more detail.
\section{5. Time-Dependent Perturbation Theory}
Now we examine the time evolution of the total system. We assume that the total initial state is
\begin{equation}
\ket{\Psi(0)}=\ket{L}\ket{0}_{A}\ket{0}_{B},
\end{equation}
The state of the total system at time $t$, expanded in terms of the system basis, is obtained as
\begin{equation}
\ket{\Psi(t)}=\sum_{n}e^{-iE_{n}t/h}|n\rangle|\chi_{n}(t)\rangle,
\end{equation}
with environmental expansion coefficients 
\begin{equation}
|\chi_{n}(t)\rangle=\exp\Big(-\frac{i\sum_{\mathcal{R}}H_{\mathcal{R}}t}{h}\Big)\langle n|U(t)|\Psi(0)\rangle,
\end{equation}
where $U(t)$ is the time evolution operator in the interaction picture. The problem is thus reduced to the evaluation of matrix elements of $U(t)$, which we have calculated in the Appendix.\\
\indent We suppose that the initial state of the system is the left-handed state $|L\rangle$. The evolved state of the total system in the localized basis of the system is written as
\begin{equation}
\ket{\Psi(t)}=\frac{1}{\sqrt{2}}\Big(\ket{\chi_{1}(t)}-\ket{\chi_{2}(t)}\Big)\ket{L}+\frac{1}{\sqrt{2}}\Big(\ket{\chi_{1}(t)}+\ket{\chi_{2}(t)}\Big)\ket{R}.
\end{equation}
We are interested in the probability of finding the system in the right-handed state, i.e.,
\begin{equation}\label{PR}
P_{R}(t)=\vert\langle R\vert\Psi(t)\rangle\vert^{2}=\frac{1}{2}\Big(\langle\chi_{1}(t)\vert\chi_{1}(t)\rangle+\langle\chi_{2}(t)\vert\chi_{2}(t)\rangle+Re\big[\langle\chi_{1}(t)\vert\chi_{2}(t)\rangle\big]\Big).
\end{equation}
\section{6. Estimation of Parameters}
To examine the dynamics of the open macroscopic quantum system, we first estimate the parameters relevant to our analysis. We start with the parameter of the system. In the semi-classical approximation, the quantum tunneling can be examined using the instanton method~\cite{Wie,Garg}, estimating the tunneling strength for a double-well potential as
\begin{equation}\label{DT}
\Delta=\frac{\Omega}{\pi}\Big(\frac{4\pi\Omega}{h}e^{2\xi}\Big)^{1/2}e^{-I/h},
\end{equation}
with 
\begin{align}
I&=\int_{-1}^{1}dx\big(2U(x)\big)^{1/2}\approx\frac{2\Omega}{3},\nonumber\\
\xi&=\int_{0}^{1}dx\Bigg(\frac{\Omega}{\big(2U(x)\big)^{1/2}}-\frac{1}{1-x}\Bigg)\approx\log 2.
\end{align}
Up to this point, our relations were quite general and not yet explicitly restricted to any particular two-level system. The parameter $h$ quantifies the macroscopicity of the system in question. The system in which $h<1$ is called a quasi-classical system. The macroscopic systems are supposed to satisfy this condition. Here, we estimate the value of $h$ for a macroscopic superconducting qubit. Such systems have been extensively studied and tested for quantum information processing (for a recent review see~\cite{Wen}). The high reproducibility of device parameters, anharmonicity of the energy level spacings and compact physical sizes make their experimental implementations feasible (see e.g.~\cite{DevS}.) The system can be basically envisaged as a superconducting ring including a Josephson junction. The electrons can flow through the junction via quantum tunneling. To manipulate the state of the system, the ring is subjected to an external magnetic field. The current flowing along the ring induces a magnetic flux $\mathsf{\Phi}$ threading through the ring with a Hamiltonian of form
\begin{equation}
\mathsf{H_{\Phi}}=\frac{\mathsf{P}^{2}}{2\mathsf{C}}+\mathsf{I_{c}}\mathsf{\Phi_{q}}U(\theta),
\end{equation}
with dimensionless potential
\begin{equation}\label{US}
U(\theta)=\frac{\gamma}{2}(\theta-\theta_{ex})^{2}-\cos\theta,
\end{equation}
in which we define
\begin{equation}
\theta=\frac{2\pi\mathsf{\Phi}}{\mathsf{\Phi_{q}}},\quad\quad \theta_{ex}=\frac{2\pi\mathsf{\Phi_{ex}}}{\mathsf{\Phi_{q}}},\quad\quad \gamma=\frac{\mathsf{\Phi_{q}}}{2\pi\mathsf{L}\mathsf{I_{c}}},
\end{equation}
where $\mathsf{C}$ is the effective electric capacitance of the junction, $\mathsf{L}$ is the self-inductance of the ring, $\mathsf{I_{c}}$ is a constant current depending on the details of the junction, $\mathsf{\Phi_{q}}=\pi\hbar/\mathsf{e}$ is the flux quantum and $\mathsf{\Phi_{ex}}$ is the flux due to the external magnetic field. If we adjust the parameters of the system as $\theta_{ex}=\pi$ and $\gamma<1$, the neighborhood of $\theta=\pi$ constitutes a symmetric double-well. This suggests that the distance in the space of flux (equivalent to $\mathsf{R}_{0}$ in the second section), is identified as $\mathsf{\Phi}_{0}=(\mathsf{\Phi_{q}}/2\pi)\theta_{0}$, which $\theta_{0}$ is the half distance between two minima (see Fig. 1). Likewise, $\mathsf{C}$ corresponds to $\mathsf{M}$. Accordingly, we find $T_{0}=\mathsf{\Phi}_{0}/(\mathsf{U}_{0}/\mathsf{C})^{1/2}$. So, the reduced Planck constant would be
\begin{equation}
h=\frac{\hbar}{\mathsf{U}_{0}\mathsf{T}_{0}}=\frac{2\mathsf{e}}{\theta_{0}(\mathsf{C}\mathsf{U}_{0})^{1/2}}=\frac{h_{0}}{\theta_{0}U_{0}^{1/2}},\quad h_{0}=2\Big(\frac{\mathsf{e}^{2}/\mathsf{C}}{\mathsf{I_{c}}\mathsf{\Phi_{q}}/2\pi}\Big)^{1/2}.
\end{equation}
To explicitly compute the constants $\theta_{0}$ and $U_{0}$, we expand (\ref{US}) up to the second order with respect to $\theta-\pi$ to find
\begin{equation}
\theta_{0}\simeq \big[6(1-\gamma)\big]^{1/2},\quad\quad U_{0}\simeq\frac{3}{2}(1-\gamma)^{2}.
\end{equation}
and accordingly,
\begin{equation}
h\simeq\frac{h_{0}}{3(1-\gamma)^{3/2}}.
\end{equation}
Let $N$ be the number of electrons constituting the junction, then one can estimate $\mathsf{C}=\mathcal{O}(N)$ and $\mathsf{I_{C}}=\mathcal{O}(N)$, and thereby $h_{0}=\mathcal{O}(N^{-1})$. By adjusting the constant $\gamma$, a macroscopic superconducting qubit with $h$ of the order of $0.1$ is realized. With this value of $h$, the tunneling strength in (\ref{DT}) is estimated as $0.001$.\\
\begin{figure}[H]
  \includegraphics[scale=0.32]{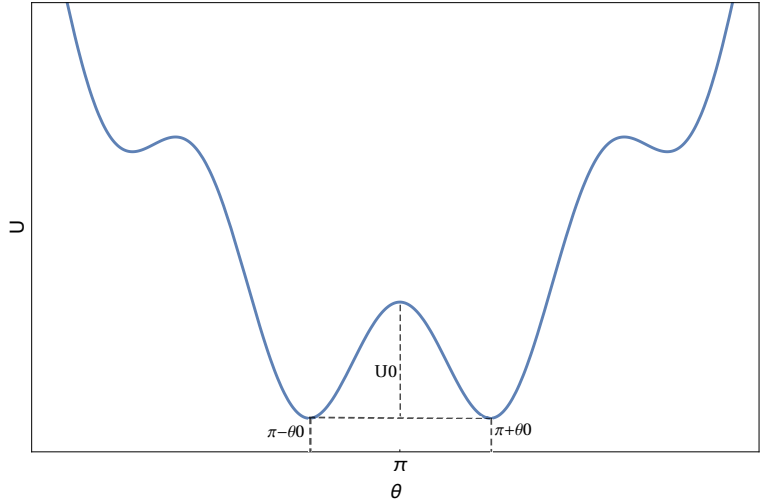}\centering
  \caption{The double well potential in terms of magnetic flux for macroscopic superconducting qubit.}
\end{figure}
The properties of the reservoir $\mathcal{R}$ are projected in the parameters of the corresponding spectral density: type parameter $s_{\mathcal{R}}$, coupling strength $J_{\mathcal{R}}$ and cut-off strength $\Lambda_{\mathcal{R}}$. The environment surrounding a superconducting system can be any type of reservoir. Since we have employed the perturbation theory, the system should be weakly coupled to the reservoirs, i.e. $J_{\mathcal{R}}\ll1$. Also, in the previous section (see Appendix), we assumed that the reservoirs are able to resolve the system's states, i.e. $\Lambda_{\mathcal{R}}\gg1$. Here, we assume that environmental parameters can be controlled by the experimenter according to aforementioned assumptions. Regarding the experimental feasibility of our setup, a number of works has been done on superconducting qubits in the context of heat transport~\cite{Gia,Gol}. The superconducting devices are susceptible to charge noise. In this context, our setup can be realized by coupling another, yet controllable reservoir to the device. The non-additive effect is emerged by adjusting the reservoir type of the controllable reservoir.
\section{7. Results and Discussion}
For an isolated quantum system, the tunneling process, according to (\ref{PI}), is manifested by symmetrical oscillations between localized states of the system (FIG. 2, orange line). Since the system is isolated, such oscillations are considered as the signature of the quantum coherence in the system. A quantum system, especially a macroscopic one, is not actually isolated. An equilibrium environment destroys the quantum coherence between the preferred states of the system. This so-called decoherence process is manifested in the reduction of the amplitude of oscillations, resulting an equilibrium steady state at long times~\cite{Gha}. \\
\indent Let us first examine the stationary case, where the reservoirs do not have dynamics. We plotted the dynamics of right-handed probability for the corresponding state, (\ref{kin}), for both identical and different reservoirs in Fig. 2, green plots. According to the plots, the system decoheres fast, regardless of the reservoirs being identical or different. Thus, at the level of kinematics, the dynamics with respect to the reservoirs is additive. \\
\indent Now we turn into the more accurate, non-stationary case, where the reservoirs also evolve in time. In our approach, if we couple the system to two identical reservoirs, the probability in (\ref{PR}) essentially reduces to $P_{R}\approx\frac{1}{2}(1-e^{-\Gamma_{2}t/2}\cos(\tilde{\Delta}t-\theta_{0}))$, where $\tilde{\Delta}$ is a modified tunneling strength and $\theta_{0}$ is a constant. The exponential term represents the decoherence effect of the environment, responsible for the decay of the oscillations, at a rate proportional to the relaxation rate of the excited state, $\Gamma_{2}$. For our macroscopic system, since the coherent evolution, characterized by $\tilde{\Delta}$, is relatively slow, the system decoheres before an oscillation is completed (FIG. 2-a, blue line).\\
\indent Now we examine the case in which the macroscopic system interacts with two different reservoirs. Such a non-equilibrium environment can be realized by modifying the parameters of the spectral densities of two reservoirs. Unlike modifying coupling strengths, which results a single equilibrium environment, modifying reservoir types and cut-off strengths comprise non-equilibrium environments. We consider the case in which the non-equilibrium environment is composed of two reservoirs of different types, so that the addition of two types would be the same as that of two identical reservoirs. The dynamics, plotted in blue line of FIG. 2-b,  shows that the non-equilibrium environment intensifies both coherent and incoherent parts of the evolution, the former is realized as several oscillations, while the latter is represented by an shorter decoherence time. This demonstrates that the non-equilibrium environment can induce coherence in the system, which by itself, being macroscopic, has a slow coherent evolution. But this comes at the expense of a faster decoherence process.\\
\begin{figure}[H]
  \subfigure[]{\includegraphics[scale=0.29]{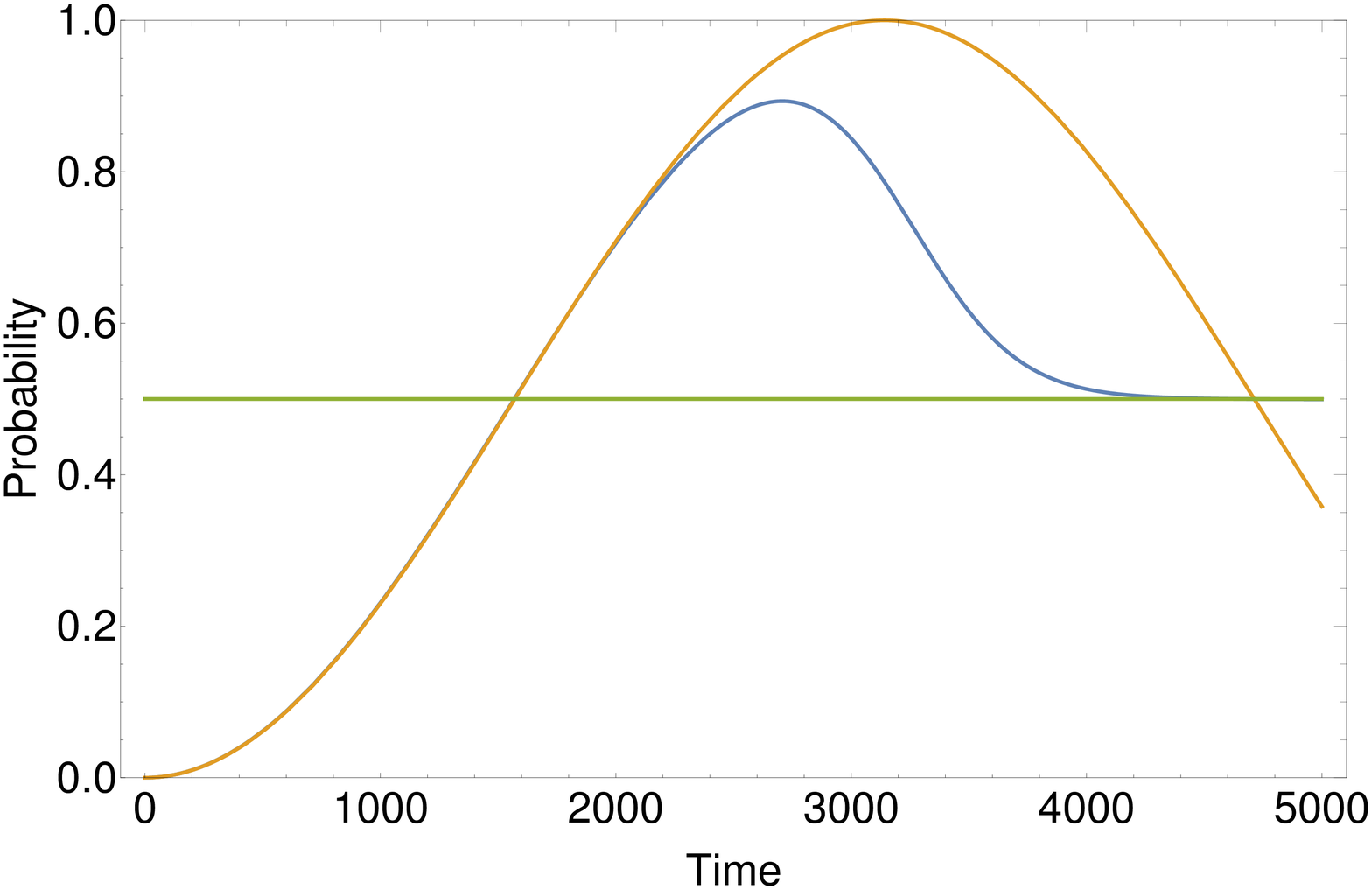}}\centering
  ~
  \subfigure[]{\includegraphics[scale=0.29]{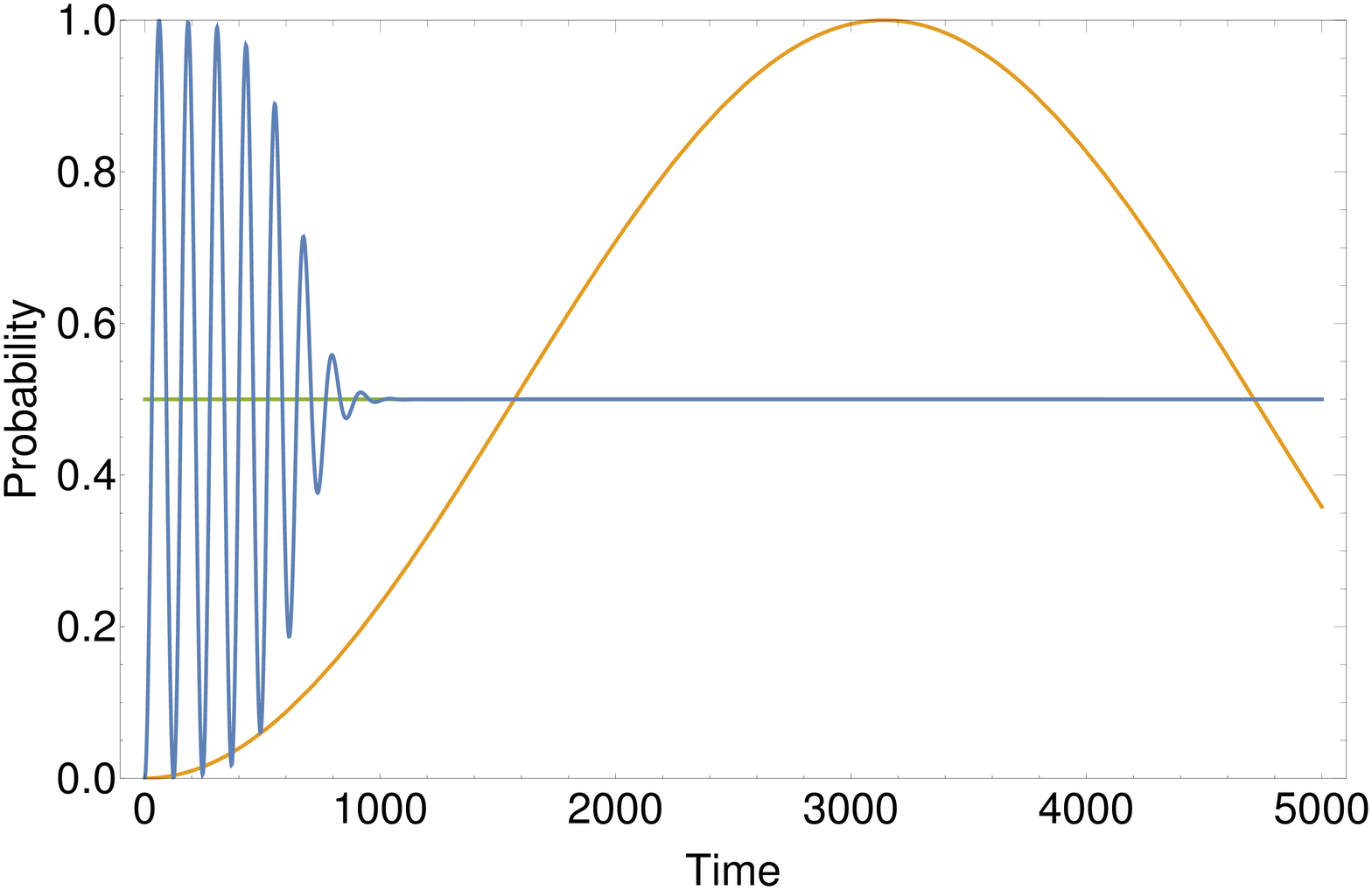}}\centering
  \caption{The dynamics of the probability of the right-handed state of the macroscopic superconducting qubit with $\Delta=0.001$. The orange line in both plots refers to the isolated system. (a) green line: with two identical ohmic stationary reservoirs, blue line: with two identical ohmic non-stationary reservoirs; (b) green line: with sub-ohmic ($s=1/2$) and super-ohmic ($s=3/2$) stationary reservoirs, blue line: with sub-ohmic ($s=1/2$) and super-ohmic ($s=3/2$) non-stationary reservoirs. The parameters of the reservoirs are $J=10^{-4}$ and $\Lambda=10$.} 
\end{figure}
\noindent Our result clearly demonstrates that when the system couples to different reservoirs, their effects are non-additive. i.e. the dynamics is affected by the interference between two reservoirs. The non-additivity of the reservoirs' effects is a signature of the non-Markovian property of the dynamics. This can be easily verified by examining the two-time environmental correlation function
\begin{equation}
C(t,t')=\sum_{n}e^{-iE_{n}(t-t')/h}\langle\chi_{n}(t')|\chi_{n}(t)\rangle,
\end{equation}
where we defined the environmental state $|\chi_{n}(t)\rangle$ in (\ref{ST}). For two identical reservoirs, the correlation function decays exponentially with time (FIG. 3-a), which is a signature of the Markovian property of the dynamics. For two different reservoirs, however, the correlation function decays with oscillations, which reflects the non-Markovian property of the environment (FIG. 3-b). This is due to the breakdown of Born approximation in our case. More precisely, since the reservoirs become entangled with the system, the interference between them can be projected in the system dynamics.
\begin{figure}[H]
  \subfigure[]{\includegraphics[scale=0.554]{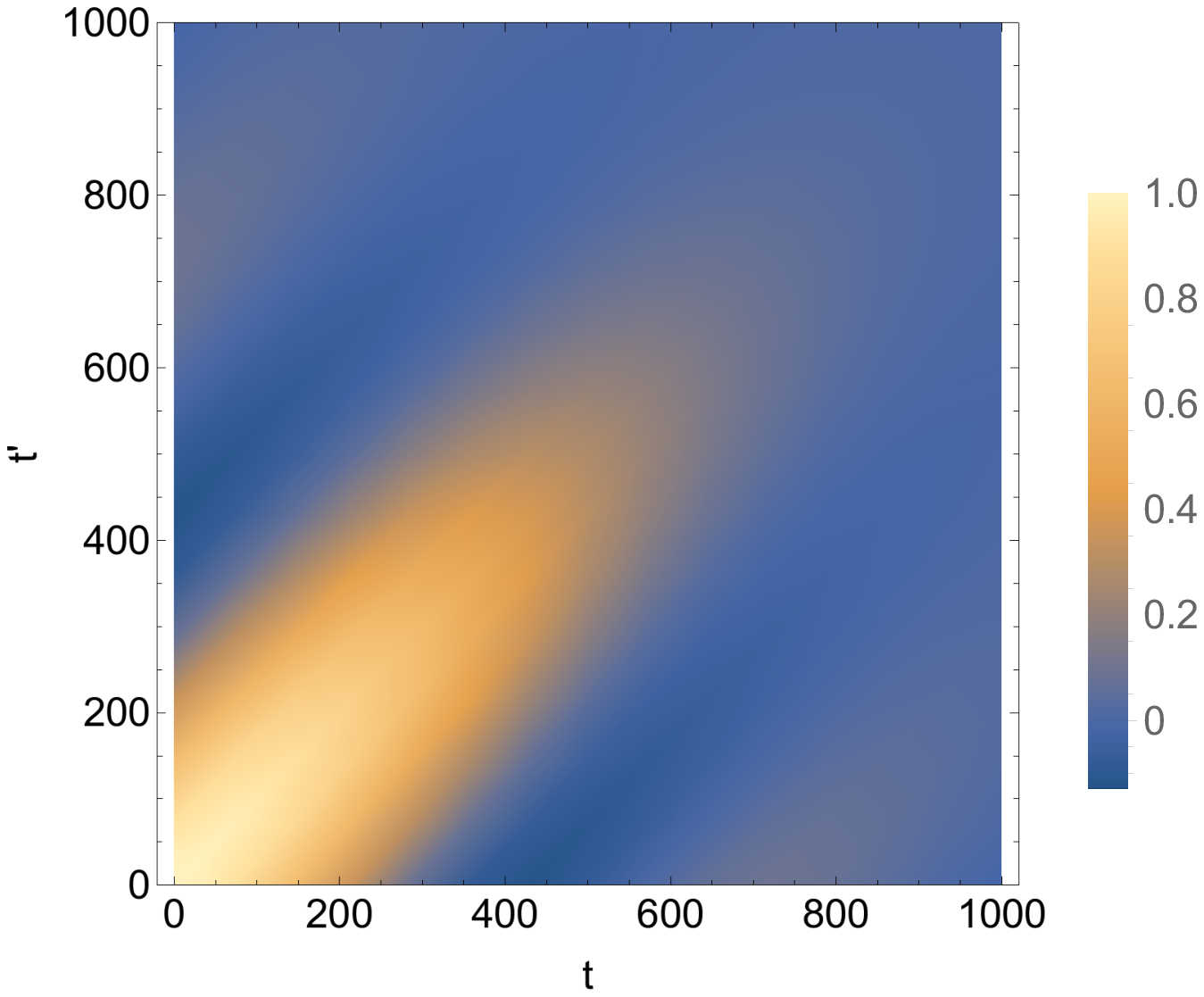}}\centering
      \hspace{1cm}
  \subfigure[]{\includegraphics[scale=0.546]{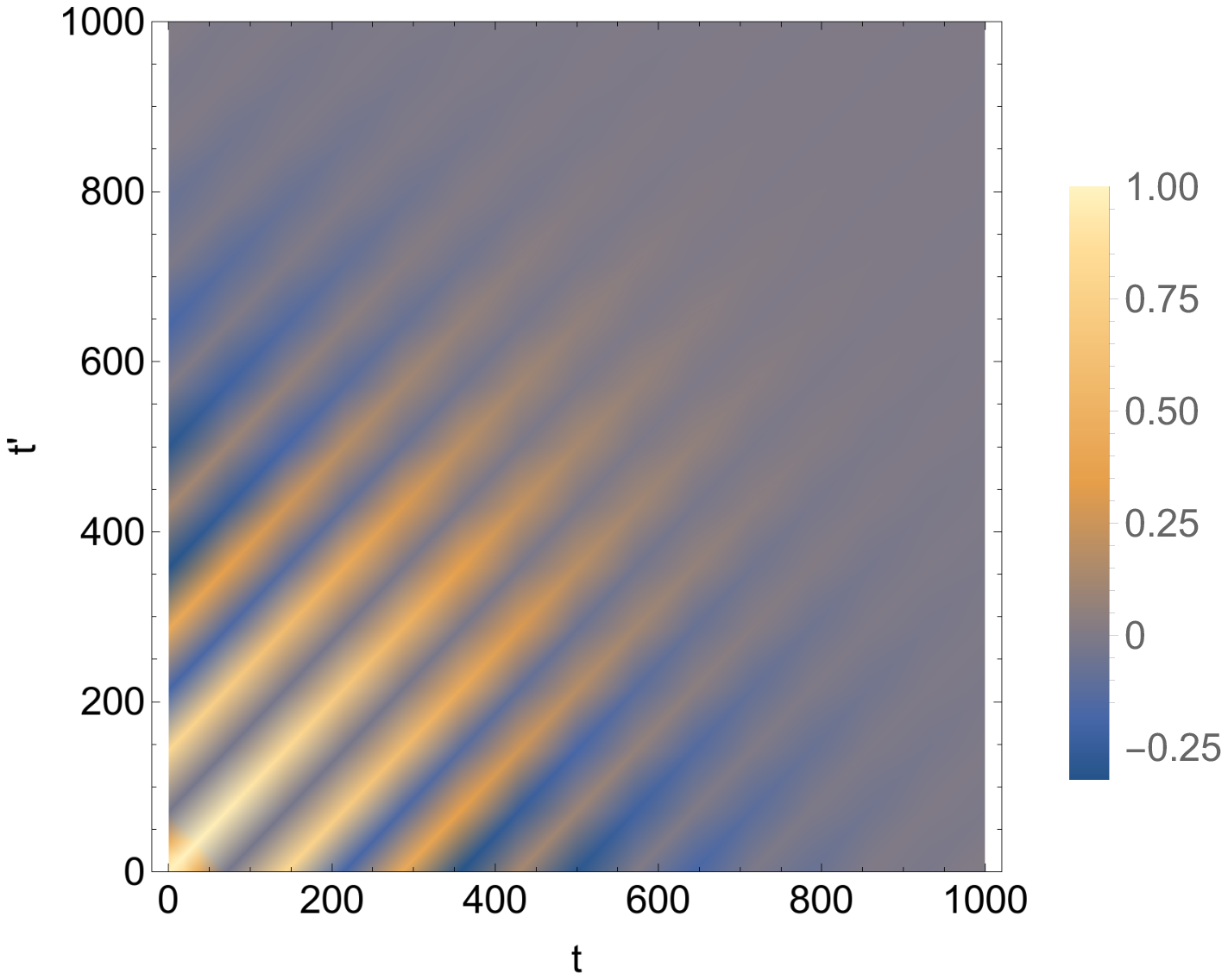}}\centering
  \caption{The two-time environmental correlation function a) for two identical reservoirs b) for two different reservoirs, depicted with the same parameters as FIG. 2.}
\end{figure}
\noindent The above analysis is based on a particular combination of two reservoirs (see FIG. 1). We can generalize it by examining all combinations in which the types of two reservoirs changes but the overall type is constant, as depicted in FIG. 4. The plot clearly shows that the non-Markovian property increases with the difference between two reservoirs.
\begin{figure}[H]
  \includegraphics[scale=0.8]{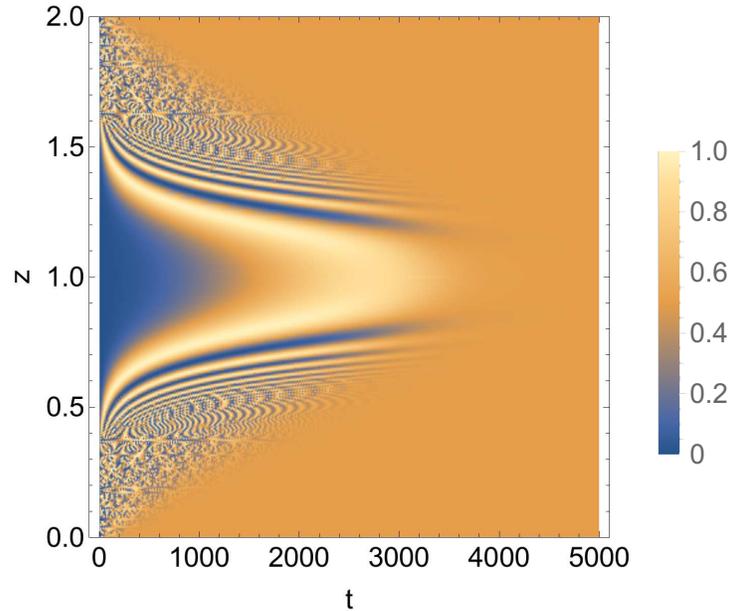}\centering
  \caption{The probability of right-handed state versus time $t$ and difference between reservoirs' type parameters $z$. The system and reservoirs' parameters are same as FIG. 2.}
\end{figure}
Our results are connected to previous works as follows. At the weak-coupling limit, three convenient assumptions, relevant to our analysis, are (I) the environment is stationary (first Born assumption), (II) the system- environment state is uncorrelated (second Born assumption), (III) the memory effects of the environment are negligible (Markov assumption). The resulting dynamical equation, known as Born-Markov master equation, predicts the additive dynamics~\cite{Pur}. Let's examine the role of the aforementioned assumptions one by one. If assumption (III) at least for one of the reservoirs is relaxed, the dynamics is non-additive~\cite{Chan}. In our work, both reservoirs are independently Markovian, ruling out assumption (III) as the origin of the non-additive dynamics. On the other hand, if assumptions (II) and (III) are relaxed, the dynamics is additive~\cite{Bra}. It can be shown that assumption (I) by itself does not lead to the non-additive dynamics (because single-reservoir correlation functions can be always set to zero by renormalizing the Hamiltonians)~\cite{Bra}. Therefor, the main source of the non-additive dynamic in our case is the entanglement of system-reservoir states, i.e. violation of the second Born assumption. Such entanglement intensifies the correlations between two reservoirs and thus increases the effect on coherent and incoherent parts, as it can be seen in Fig. 2.\\
\indent We benchmarked our approximate result to exact result of Mitchison and Plenio~\cite{Mit}. They have shown that at long times, when the system is locally coupled to different reservoirs, the non-additive effect is observed. In our work, however, only short-time dynamics is affected by the interference between two reservoirs. The difference is due to the fact that we couple the system globally to the reservoirs, while they couple different regions of the system locally to different reservoirs, thus there would be currents between different regions of the system. Such currents are the source of long-time non-additive effects. These results are also confirmed by numerical analysis of two coupled damped oscillators~\cite{RiP}. They showed that (see Fig. 8 of~\cite{RiP}) even at very weak coupling limit, the non-additive contribution accumulates with time, leading to long-time currents.
\section{Conclusion}
We examined the dynamics of a macroscopic quantum system, i.e. a macroscopic superconducting qubit, weakly coupled to two different, independent vacuum reservoirs. We employed the second-order, time-dependent perturbation theory to capture the accurate dynamics. It has been shown that for a composite system, coupled locally to different reservoirs, the overall effect is non-additive i.e. the dynamics is affected by the interference between the reservoirs. Our results show that, when the system interacts globally with two different reservoirs, the short-term dynamics is also non-additive. The non-additive term generates coherences but intensifies the decoherence process. The source of this effect in our work is traced back to the  entanglement of the states of the system and the reservoirs. Such a setup can be employed to induce coherence in the macroscopic quantum systems at the price of shorter decoherence times.\\

\appendix*
\section{Appendix: Calculation of Matrix Elements of $U_{I}(t)$}
\setcounter{equation}{0}
We expand $U(t)$ up to the second order with respect to the interaction Hamiltonian $H_{int}(t)=\sum_{\mathcal{R}}H_{S\mathcal{R}}(t)$ as
\begin{equation}\label{11}
U(t)\simeq 1-\frac{i}{h}\int_{0}^{t}dt_{1}H_{int}(t_{1})-\frac{1}{h^{2}}\int_{0}^{t}dt_{2}\int_{0}^{t_{2}}dt_{1} H_{int}(t_{2}) H_{int}(t_{1})
\end{equation}
\noindent We first calculate
\begin{equation}
U(t)\ket{0}_{A}\ket{0}_{B}\simeq U_{0A,0B}(t)\ket{0}_{A}\ket{0}_{B}+\sum_{\alpha}\Big\{U_{\alpha A,\alpha B}(t)\ket{\alpha}_{A}\ket{\alpha}_{B}+U_{0A,\alpha B}(t)\ket{0}_{A}\ket{\alpha}_{B}+U_{\alpha A,0B}(t)\ket{\alpha}_{A}\ket{0}_{B}\Big\},
\end{equation}
where
\begin{align}\label{Us}
U_{0A,0B}(t)=&1-\frac{i }{2h}\sum_{\mathcal{R}}\sum_{\alpha}\int_{0}^{t}dt_{1}f_{\alpha,\mathcal{R}}^{2}(t_{1})-\frac{1}{2h}\sum_{\mathcal{R}}\sum_{\alpha}\omega_{\alpha,\mathcal{R}}
\int_{0}^{t}dt_{2}\int_{0}^{t_{2}}dt_{1}e^{-i(t_{2}-t_{1})\omega_{\alpha,\mathcal{R}}}f_{\alpha,\mathcal{R}}(t_{2})f_{\alpha,\mathcal{R}}(t_{1})\nonumber\\
&~-\frac{1}{4h^{2}}\sum_{\alpha}\int_{0}^{t}dt_{2}\int_{0}^{t_{2}}dt_{1}\Big\{f_{\alpha,A}^{2}(t_{2}),f_{\alpha,B}^{2}(t_{1})\Big\}\nonumber\\
U_{\alpha A,\alpha B}(t)=&-\frac{1}{2h}\omega_{\alpha,A}^{1/2}\omega_{\alpha,B}^{1/2}\int_{0}^{t}dt_{2}\int_{0}^{t_{2}}dt_{1}
\Big\{e^{i\omega_{\alpha,A}t_{2}}f_{\alpha,A}(t_{2}),e^{i\omega_{\alpha,B}t_{1}}f_{\alpha,B}(t_{1})\Big\}\nonumber\\
U_{0A,\alpha B}(t)=&\frac{i}{\sqrt{2h}}\omega_{\alpha,B}^{1/2}\int_{0}^{t}\!dt_{1}e^{i\omega_{\alpha,B}t_{1}}f_{\alpha,B}(t_{2})
+\frac{1}{\sqrt{8h^{3}}}\omega_{\alpha,B}^{1/2}\int_{0}^{t}dt_{2}\!\int_{0}^{t_{2}}dt_{1}
\Big\{f_{\alpha,A}^{2}(t_{2}),e^{i\omega_{\alpha,B}t_{1}}f_{\alpha,B}(t_{1})\Big\},
\end{align}
to find
\begin{align}\label{ST}
\ket{\chi_{n}(t)}=&\ket{0}_{A}\ket{0}_{B}\bra{n}U_{0A,0B}(t)\ket{Li}+\sum_{\alpha}e^{-i\omega_{\alpha,B}t}\ket{0}_{A}\ket{\alpha}_{B}\bra{n}U_{0A,\alpha B}(t)\ket{L}\nonumber\\
&+\sum_{\alpha}e^{-i\omega_{\alpha,A}t}\ket{\alpha}_{A}\ket{0}_{B}\bra{n}U_{\alpha A,0B}(t)\ket{L}+\sum_{\alpha}e^{-i(\omega_{\alpha,A}+\omega_{\alpha,B})t}\ket{\alpha}_{A}\ket{\alpha}_{B}\bra{n}U_{\alpha A,\alpha B}(t)\ket{L},
\end{align}
where we define $f_{\alpha,\mathcal{R}}(t)=\omega_{\alpha,\mathcal{R}}f_{\alpha,\mathcal{R}}(x(t))$,  $x(t)=e^{iH_{S}t/h}xe^{-iH_{S}t/h}$ and $\{g(t),h(t')\}=g(t)h(t')+g(t')h(t)$. Note that $U_{\alpha A,0B}(t)$ can be obtained by interconverting $A\leftrightarrow B$ of $U_{0A,\alpha B}(t)$ in (\ref{Us}).\\
\indent The problem is now reduced to the evaluation of matrix elements of interaction Hamiltonian in (\ref{ST}). The potential $U(x)$ being an even function, the energy eigenstates $\ket{n}$ have definite parity. Since $U_{0A,0B}(t)$ and $U_{\alpha A,\alpha B}(t)$ are even functions and $U_{0A,\alpha B}(t)$ and $U_{\alpha A,0B}(t)$ are odd functions, the following selection rules are identified
\begin{align}\label{15}
\bra m U_{0A,0B}\ket{n}&=\bra m U_{\alpha A,\alpha B}\ket{n}=0, \nonumber\\
\bra n U_{0A,\alpha B}\ket{n}&=\bra n U_{\alpha A,0B}\ket{n}=0.
\end{align}
The diagonal matrix elements of $U_{0A,0B}(t)$ are evaluated as
\begin{align}\label{U00}
\bra{n}U_{0A,0B}(t)\ket{n}=1&-\frac{it}{h}\sum_{\mathcal{R}}\Bigg\{\delta E_{n,\mathcal{R}}^{(1)}-\frac{1}{\pi}\sum_{m}x_{mn}^{2}\int_{0}^{\infty}d\omega_{\mathcal{R}}
\frac{J(\omega_{\mathcal{R}})}{\omega_{\mathcal{R}}+\Omega_{mn}}\Bigg\}\\\nonumber &
-\frac{1}{\pi h}\sum_{\mathcal{R}}\sum_{m}x_{mn}^{2}\int_{0}^{\infty}d\omega_{\mathcal{R}}J(\omega_{\mathcal{R}})
\frac{1-e^{\imath t(\omega_{\mathcal{R}}+\Omega_{mn)}}}{(\omega_{\mathcal{R}}+\Omega_{mn})^{2}}-\frac{t^{2}}{h^{2}}\delta E_{n,A}^{(1)}\delta E_{n,B}^{(1)},
\end{align}
where $J(\omega_{\mathcal{R}})$ is the spectral density of the reservoir $\mathcal{R}$, corresponding to a continuous spectrum of environmental frequencies, $\omega_{\mathcal{R}}$, defined as
\begin{equation}
J(\omega_{\mathcal{R}})=\frac{\pi}{2}\sum_{\alpha}\gamma_{\alpha,\mathcal{R}}^{2}\omega_{\alpha \mathcal{R}}^{3}\delta(\omega_{\mathcal{R}}-\omega_{\alpha,\mathcal{R}})\equiv J_{\mathcal{R}}\omega_{\mathcal{R}}\big(\frac{\omega_{\mathcal{R}}}{\Lambda_{\mathcal{R}}}\big)^{s_{\mathcal{R}}-1}e^{-\omega_{\mathcal{R}}/\Lambda_{\mathcal{R}}},
\end{equation}
where $s_{\mathcal{R}}$, $J_{\mathcal{R}}$ and $\Lambda_{\mathcal{R}}$ are the type parameter, coupling strength and cut-off strength of reservoir $\mathcal{R}$. The different types of reservoirs are characterized by the value of parameter $s_{\mathcal{R}}$ as sub-ohmic ($0<s_{\mathcal{R}}<1$), ohmic ($s_{\mathcal{R}}=1$) and super-ohmic ($s_{\mathcal{R}}>1$). For the equilibrium environment, we consider each reservoir to be ohmic. \\
\noindent The quantity embraced by the bracket on the right-hand side of (\ref{U00}) coincides with $\delta E_{n,\mathcal{R}}$ in (\ref{DE}). Thus, the following expression is valid up to the second order
\begin{align}\label{U00n}
\bra{n}U_{0A,0B}(t)\ket{n}\simeq e^{-it/h\sum_{\mathcal{R}}\delta E_{n,\mathcal{R}}}&\Bigg\{1-\frac{1}{\pi h}\sum_{\mathcal{R}}\sum_{m}x_{mn}^{2}\int_{0}^{\infty}d\omega_{\mathcal{R}}J(\omega_{\mathcal{R}})\nonumber\\
&\times\Big(\frac{2\sin^{2}\{(\omega_{\mathcal{R}}+\Omega_{mn})t/2\}}{(\omega_{\mathcal{R}}+\Omega_{mn})^{2}}-i\frac{\sin\{(\omega_{\mathcal{R}}
+\Omega_{mn})t\}}{(\omega_{\mathcal{R}}+\Omega_{mn})^{2}}-\frac{t^{2}}{h^{2}}\delta E_{n,A}^{(1)}\delta E_{n,B}^{(1)}\Big)\Bigg\}.
\end{align}
\noindent We assume that the reservoir's cut-off strength is much higher than the system's characteristic strength (Markov approximation), i.e. $\Lambda_{\mathcal{R}}\gg\Omega$, so that at times much higher than $\Omega^{-1}$, the first term of the integral in (\ref{U00n}) can be approximated by a delta function $\delta(\omega_{\mathcal{R}}+\Omega_{mn})$. Obviously, the result of the corresponding integral would be $J(\omega_{\mathcal{R}}+\Omega_{mn})$, which is zero for $\Omega_{mn}\geq0$. The elements of $U_{0A,0B}(t)$ are then reduced to
\begin{align}
\bra{1}U_{0A,0B}(t)\ket{1}&\simeq e^{-it/h\sum_{\mathcal{R}}\delta E_{1,\mathcal{R}}}\Bigg\{1+\frac{i}{\pi h}\sum_{\mathcal{R}}\int_{0}^{\infty}d\omega_{\mathcal{R}}J(\omega_{\mathcal{R}})\frac{\sin\{(\omega_{\mathcal{R}}
+\Delta)t\}}{(\omega_{\mathcal{R}}+\Delta)^{2}}-\frac{t^{2}}{h^{2}}\delta E_{1A}^{(1)}\delta E_{1B}^{(1)}\Bigg\},\nonumber\\
\bra{2}U_{0A,0B}(t)\ket{2}&\simeq e^{-it/h\sum_{\mathcal{R}}\delta E_{2,\mathcal{R}}}\Bigg\{\!1-\!\sum_{\mathcal{R}}\frac{\Gamma_{2,\mathcal{R}}t}{2}+\frac{i}{\pi h}\sum_{\mathcal{R}}\int_{0}^{\infty}d\omega_{\mathcal{R}}J(\omega_{\mathcal{R}})\frac{\sin\{(\omega_{\mathcal{R}}
-\Delta)t\}}{(\omega_{\mathcal{R}}-\Delta)^{2}}-\frac{t^{2}}{h^{2}}\delta E_{2A}^{(1)}\delta E_{2B}^{(1)}\Bigg\}.
\end{align}
If the coupling between the system and each reservoir is considered to be weak, we have $\Gamma_{2,\mathcal{R}}\ll\Omega$. At the temporal domain $\Omega^{-1}\ll t\ll\Gamma_{2,\mathcal{R}}^{-1}$, we finally obtain
\begin{align}\label{U00f}
\bra{1}U_{0A,0B}(t)\ket{1}&\approx\prod_{\mathcal{R}}\exp\Bigg\{-\frac{i}{h}\Big(t\delta E_{1,\mathcal{R}}-\frac{1}{\pi}\int_{0}^{\infty}d\omega_{\mathcal{R}}J(\omega_{\mathcal{R}})\frac{\sin\{(\omega_{\mathcal{R}}
+\Delta)t\}}{(\omega_{\mathcal{R}}+\Delta)^{2}}\Big)-\frac{t^{2}}{h^{2}}\delta E_{1A}^{(1)}\delta E_{1B}^{(1)}\Bigg\},\nonumber\\
\bra{2}U_{0A,0B (t)}\ket{2}&\approx\prod_{\mathcal{R}}\exp\Bigg\{\!\!-\frac{\Gamma_{2,\mathcal{R}}t}{2}-\frac{i}{h}\Big(t\delta E_{2,\mathcal{R}}-\frac{1}{\pi}\int_{0}^{\infty}d\omega_{\mathcal{R}}J(\omega_{\mathcal{R}})\frac{\sin\{(\omega_{\mathcal{R}}
-\Delta)t\}}{(\omega_{\mathcal{R}}-\Delta)^{2}}\Big)-\frac{t^{2}}{h^{2}}\delta E_{2A}^{(1)}\delta E_{2B}^{(1)}\Bigg\}.
\end{align}
The elements of $U_{\alpha A,\alpha B}(t)$ are obtained as
\begin{align}
\bra{n}U_{\alpha A,\alpha B}(t)\ket{n}=&-\frac{\pi}{4h}\gamma_{\alpha,A}\omega_{\alpha,A}^{3/2}\gamma_{\alpha,B}\omega_{\alpha,B}^{3/2}\nonumber\\
&\times\Bigg\{\frac{1-e^{-i(\omega_{\alpha,A}+\omega_{\alpha,B})t}}{(\omega_{\alpha,A}+\omega_{\alpha,B})(\omega_{\alpha,B}+(-1)^{n+1}\Delta)}+\frac{e^{-i(\omega_{\alpha,A}+(-1)^{n+1}\Delta)t}-1}{(\omega_{\alpha,A}+(-1)^{n+1}\Delta)(\omega_{\alpha,B}+(-1)^{n+1}\Delta)}\nonumber\\
&+\frac{1-e^{-i(\omega_{\alpha,A}+\omega_{\alpha,B})t}}{(\omega_{\alpha,A}+\omega_{\alpha,B})(\omega_{\alpha,A}+(-1)^{n+1}\Delta)}+\frac{e^{-i(\omega_{\alpha,B}+(-1)^{n+1}\Delta)t}-1}{(\omega_{\alpha,A}+(-1)^{n+1}\Delta)(\omega_{\alpha,B}+(-1)^{n+1}\Delta)}\Bigg\}.
\end{align}
The elements of $U_{0A,\alpha B}(t)$ are evaluated as
\begin{equation}\label{U0A}
\bra{m}U_{0A,\alpha B}(t)\ket{n}=\sqrt{\frac{\pi}{h}}\Big(i+\frac{t}{h}\delta E_{n,A}^{(1)}\Big)\gamma_{\alpha,B}^{2}\omega_{\alpha,B}^{3}\frac{\sin\big\{(\omega_{\alpha,B}+(-1)^{m}\Delta)t/2\big\}}{\omega_{\alpha,B}+(-1)^{m}\Delta}
e^{i(\omega_{\alpha,B}+(-1)^{m}\Delta)t/2}.
\end{equation}
The elements of $U_{\alpha A,0B}(t)$ are calculated by interconverting $A\leftrightarrow B$ in $\bra{m}U_{0A,\alpha B}(t)\ket{n}$ in (\ref{U0A}).\\

\end{document}